\documentclass[preprint,amsmath,amssymb]{revtex4}

\usepackage{graphicx}
\usepackage{dcolumn}
\usepackage{bm}
\usepackage{verbatim}

\begin{document}

\preprint{}

\title{ Phase Separation in the Wake of Moving Fronts:
Experiments and Simulations}

\author{P\'eter Hantz} 
\altaffiliation[Also at the ]{ Department of Plant Taxonomy and Ecology, E\"otv\"os Lor\'and University,
P\'azm\'any s\'et\'any 1/C 1117 Budapest (Hungary) }
\author{Istv\'an Biro}
\affiliation{Department of Theoretical Physics, Babe\c{s}--Bolyai University,
Str.~Kog\u{a}lniceanu Nr.~1/B, 400084 Cluj (Kolozsv\'ar),  (Romania) }

\date{\today}

\begin{abstract}

The formation of regular precipitate stripes in the wake of moving 
chemical reaction-diffusion fronts is investigated. Experiments on the $NaOH+CuCl_2$
reaction in PVA hydrogel yield stripes parallel or slightly oblique to the 
front that supplies the precursor of the precipitate. 
The pattern formation was modeled by phase separation described by the Cahn-Hilliard Equation. 
Computer simulations reproduced the parallel and the oblique striping as well. 
Stripes perpendicular to the front are unstable and cannot be observed, in complete agreement 
with the experiments.  

Pattern formation in the wake of quenching fronts has also been investigated 
computationally, and compared to the previous results. It has been shown that below a 
certain front speed stripes perpendicular to the front will appear. Moreover, 
they will bend so that their growing end to be kept perpendicular even if the front 
changes its direction. This result can be important in designing several nanotechnological and 
lithographical processes. 
\end{abstract}

\maketitle

\section{Introduction }
Recently, there is growing interest in spontaneous pattern forming processes that can 
yield regular microscopic textures. A widely studied example is phase separation,
shown by various chemical and physical 
processes \cite{{gen1},{gen2},{gen3},{gen4},{gen5},{cheq}}. 
Most of the experiments  are concerned with initially homogeneous systems, where pattern formation 
starts after a temperature quench shifts the system into an unstable state. 
However, a new stream of research is being defined by studies on those processes where phase 
separation takes place in the wake of moving fronts. 

Two mechanisms are known to yield spinodal phase separation behind traveling fronts. 
In the first, the concentration of the phase-separating compound lies between the spinodal points, 
but the temperature only drops below the critical value, required for the instability to occur, 
behind a quenching front. Alternatively, when the concentration is initially in the stable 
regime, a source front shifting it in between the spinodal points can switch the system 
into the unstable, pattern-forming range. Some important aspects on these mechanisms are listed
as follows. 

Computational studies on phase separation under directional quenching are presented
in \cite{furukawa} and \cite{liu}. The phase separation was studied 
is the framework of the Cahn-Hilliard equation. At high velocities of the cooling front, 
irregular morphology (IM) emerged. At  decreasing front velocities, stripes parallel to the 
quenching front, termed lamellar morphology (LM), and stripes perpendicular to the front, termed 
columnar morphology (CM) were found. Although little attention has been paid to the systematic 
investigation of the textures when the stripes were oblique to the front, we will consider 
these kind of patterns, and refer to them as oblique morphology (OM).    

The examination of phase separation in the wake of source fronts has been motivated by the desire to
set up a minimal model of the Liesegang phenomenon, in which a series of
precipitate stripes emerge in the wake of the diffusion front of a reagent, referred to as the 
``outer electrolyte'', that penetrates into a hydrogel containing 
an ``inner electrolyte'' \cite{{liese},{fialkowski}}. 
Modeling the formation of the one-dimensional Liesegang patterns has been achieved by assuming that the reaction 
of the electrolytes yields an intermediary compound first, that separates into high and low density regions
according to the Cahn-Hilliard equation  \cite{{chliese},{din}}.

Several experimental results have been reported on two-dimensional striped structures formed in the 
wake of filiform source fronts. Besides the classical Liesegang experiments \cite{fialkowski}, 
a great variety of such patterns have been found in the  $NaOH+CuCl_2$ reaction in polyvinyl-alcohol 
(PVA) hydrogel medium \cite{hantz2}. Parallel and oblique morphologies have also been observed in this
reaction. However, up to this time, no mathematical models have been elaborated for describing 
these formations.
  
In this paper we investigate the formation of the microscopic, striped (``secondary'') patterns 
emerging in the $NaOH+CuCl_2$ reaction running in PVA hydrogel sheets.
The main characteristics of the parallel and oblique morphologies observed in the experiments have been 
reproduced by computer simulations based on phase separation described by the Cahn-Hilliard equation.

\section{Experiment}

The $NaOH+CuCl_2$ reaction has been studied in a ``Liesegang-like'' setup, with $NaOH$ 
as the outer electrolyte, and $CuCl_2$ as the inner electrolyte homogenized in PVA
hydrogel. Details of the sample preparation are described in \cite{{hantz2},{hantz5}}. 
The $NaOH$ penetrates into the gel by diffusion, and its reaction with the inner electrolyte 
leads to a great variety of precipitate structures \cite{{hantz1},{hantz2},{hantz3},{hantz4},{hantz6}}. 
These have been classified into ``primary'' and ``secondary'' patterns \cite{hantz2}. 
In the following, the formation of the secondary, striped microscopic patterns is examined.   

First, the reaction-diffusion front of the $NaOH$ sweeps through the gel, which 
is assumed be the source of a compound that phase separates into high-density ($CuO$ precipitate) 
and low-density (free of precipitate) stripes. 
The $CuO$ precipitate stripes are mostly parallel to the reaction-diffusion front, and in some cases oblique 
morphology has also been found. Although the source front cannot be observed by the naked eye,  
the oblique morphology can be identified by an intrinsic characteristic: 
In the case of the lamellar morphology, the stripes appear one by one along the source front, that is, their 
growth does not proceed via the elongation of their end points. In contrast, in the case of the oblique 
morphology, the stripes have growing endpoints in the wake of the source front, and they form an angle with the 
envelope of their terminal points \footnote{The spiral patterns in the classical two-dimensional Liesegang 
experiments \cite{henish} can also be considered as a special case of the oblique morphology.}. 

The character of the striped microscopic patterns is correlated with the speed of the source front.
Although the source front is not visible, it is followed by a sharp precipitation front,
where a blue-green compound is formed that shows no structure when investigated with an optical microscope 
\cite{hantz2}. This is referred to as the 'active border' of the blue-green precipitate, it can easily 
be observed even by the naked eye, and is likely to follow the shape and the speed of the source front. 
If the front velocity is between $0.6-0.9\ \mu /s$, uniformly distributed colloidal precipitate 
forms ahead of the active  border, but behind the source front. When the velocity is even smaller, 
the precipitate starts 
to show a pattern: a regular structure of parallel stripes of colloidal precipitate 
appears. Smaller front velocities lead to longer stripe wavelengths \cite{hantz2}. Note that in contrast to 
the hypotheses presented in \cite{hantz2}, it is not believed any more that the active border is the source 
of the stripes' precursor.  

The first emerging stripes will be parallel to the active border. When the source front does not change its 
shape and orientation, the subsequent stripes will form parallel to the previous stripes and the front as well, 
giving rise to a lamellar morphology.
According to video microscope observations, a possible scenario for the formation of the oblique stripes is the 
following: 
When the source front suddenly changes its shape or orientation, {\it e.g.} as a result of an influx of the outer 
electrolyte from a novel direction, the newly formed stripes cannot follow the front's altered orientation, 
but form more or less parallel to the 
previous ones. The newly formed stripes elongate only up to the limit of the region already 
visited by the source front, with the envelope of their growing endpoints being parallel to the actual position of 
the front (fig. 1).  

\section{Modeling the pattern formation in the wake of source fronts}

The process of phase separation, occurring in the wake of the
$NaOH+CuCl_2$ reaction-diffusion front, has been modeled by the Cahn-Hilliard
equation with a Ginzburg-Landau free energy \cite{cheq}. Although one of the equilibrium 
densities of this free energy is negative, and therefore, in our case unphysical, 
the equation can easily be rescaled to a form where both equilibrium
densities are positive. However, for the sake of simplicity and without
offending the physical content, we have used the free energy with minima at $-1$
and $+1$.

In order to focus our attention to the pattern formation, the reaction-diffusion 
system that produces the phase separating chemical has not been included in our model. 
In order to describe the source front, a Gaussian-type source term $S(x,t; v)$ has 
been added to the Cahn-Hilliard Equation:

\begin{equation}
\frac{\partial c(x,y,t)}{\partial t}=
-\Delta[c(x,y,t)-c(x,y,t)^3+\epsilon \Delta c(x,y,t)]+S(x,y,t;v)
\label{ch}
\end{equation}
where
\begin{equation}
S(x,y,t;v)=A\cdot \exp \left[ -\alpha (ax+b-vt)^2 \right];
\label{Gauss}
\end{equation}

Initially, the concentration of the compound $c$ is set to the stable magnitude $c_0(x,y,0)=-1+\eta$
in the whole rectangular simulation area, where $\eta$ is  a random uniform
deviate distributed between $\pm 0.001$ \cite{liu}. This deviate has been added to the model 
in order to make it more realistic. 

The value $c_0$ is increased by the source, moving with constant 
speed $v$, to the constant value $c_f$ \cite{gr}. The speed $v$ of the source, as 
well as the concentration $c_f$ next to the source front, are considered as independent 
simulation parameters. 
Having the speed $v$ fixed, the value  of $c_f$ is determined by the 
amplitude $A$ and the width $\alpha$ of the Gaussian source. If $c_f$ lies
in between the spinodal points, that is, $-1/\sqrt 3<c_f<1/\sqrt 3$, 
the system will be unstable against linear perturbations, and phase separation 
will take place in the wake of the front. As time goes on, the
concentration profile $c(x,y,t)$ tends to reach the equilibrium values, and a ``ripening''
of the regions with the stable concentrations will take place as well. However, the
initial conditions, as well as the movement of the Gaussian-type source "front" will 
strongly affect the emerging patterns. These features will be our primary concern.   

The  equation (\ref{ch}) was solved on a rectangular grid using the finite difference
method \footnote{We used the nine-point Laplacian 
$\Delta c(x_i,y_j)=\frac{1}{6}[4c(x_{i-1},y_j)+4c(x_{i+1},y_j)+4c(x_i,y_{j+1})+4c(x_i,y_{j-1})
+c(x_{i-1},y_{i-1})+c(x_{i-1},y_{j+1})+c(x_{i+1},y_{j-1})+c(x_{i+1},y_{j+1})
-20c(x_i,y_j)]$ 
and the fourth order term was approximated by 
$\Delta^2 =
c(x_{i-2},y_j)+c(x_{i+2},y_j)+c(x_i,y_{j+2})+c(x_i,y_{j-2})
+2[ c(x_{i-1},y_{j-1})+c(x_{i-1},y_{j+1})+c(x_{i+1},y_{j-1})+c(x_{i+1},y_{j+1}) ]
-8[ c(x_{i-1},y_j)+c(x_{i+1},y_j)+c(x_i,y_{j+1})+c(x_i,y_{j-1}) ] +20c(x_i,y_j)
$}. 
Periodic boundary conditions have been used in both directions. 

The time evolution of the system was computed by explicit simple time
marching on a rectangular grid. The mesh size 
was $\Delta x= \Delta y =1$, while the time step was
$\Delta t=0.01$. A negligible change in some selected simulation results was only observed when 
the mesh size was halved and the time step was diminished $10$ times. 
It is also important to mention that the effect of the grid anisotropy 
on the simulation results was also of minor importance. This has been checked by 
comparing pattern formation in the wake of source fronts with different orientations.  

\section{Simulation results: source fronts traveling with constant speed}

The speed of the source front, as well as the initial conditions, play a decisive 
role in determining the character of the pattern formation. In this section, 
our concern will be to investigate their influence. 

Initially, the concentration was set to $c_0(x,y,0)=-1+\eta$ in the simulation area. 
This was increased by the front to $c_f$, a value being in between the 
spinodal points. In the following, the results for the case $c_f=0$ will be presented, 
and major differences for $c_f\ne 0$ will be mentioned. 

In our investigations, the source fronts were traveling with different constant speed values.
Although in the experiments the front speed varies in time, this change is usually not 
significant for $5-10$ stripe wavelength, and the front speed
can be considered locally constant.  

Different pre-patterns with $c(x,y,0)=0$ introduced in the $x\in(5,30)$ space units region of the 
simulation area highly affected the character of the patterns emerging even after the front 
sweeps through this region.  Note that the source was started at
$x=10$ space units from the $Y-$axis of the simulation area.

In the following, patterning at three different initial conditions are presented.
The effects of the front speeds are also discussed within these cases.

a.) In the simplest scenario, the front is started parallel to the $Y-$axis of the 
rectangular grid, and sweeps with constant speed and orientation toward the opposite edge.
The concentration is $c_0(x,y,0)=-1+\eta$ all over the simulation area, that is, no initial 
patterning is introduced in the system. Depending on the front speed, two characteristic 
morphologies were observed. 

If the front speed is higher than $v\approx 5$, an irregular morphology builds up in
the wake of the front. The explanation is straightforward: the relatively slow phase separation
drops behind the rapidly progressing front. As a consequence, there will be a large
unstable domain between the front and the region where phase separation occurs, and 
the scenario will essentially be the same as the phase separation in a field with a
homogeneous concentration in the unstable regime.

At front speeds below $v\approx 2$, stripes parallel to the front are formed. This pattern is 
referred to as a lamellar morphology. In the vicinity of the limit velocity, the stripes 
are staggered and coarse, but below $v\approx 1$, they become smooth and straight. 
Regular lamellar morphology appears at much lower front speeds as well, but the 
wavelength of the stripes is increased. This effect is reminescent of the results
encountered in modeling the Liesegang phenomenon by the terms of a spinodal phase separation,
namely the growing wavelength of the stripes in the wake of a source front with 
decreasing speed \cite{chliese}. Note that the character of the patterns does not
changed when $c_f=\pm 0.12$. 

b.) In order to examine the stability of the lamellar morphology, the simulations 
have been performed such that randomly distributed spots with  
$c(x,y,0)=0$ were introduced in the $x\in(5,30)$ space units region of the simulation area.
Despite the above random initial conditions, which disturb the first stripes that build up 
in the wake of the front, lamellar  morphology, or oblique morphology, 
with a small angle, appears at $v\in (0.5,2)$. At $v>5$ irregular morphology, 
and at $v\approx 0.1$ spotty irregular morphology appears in a $250$x$620$ simulation area. 
The tendency to form the lamellar morphology slightly diminished when $c_f=-0.12$.

c.) In certain parameter regions, the form and orientation of the preceding stripes will 
strongly influence the location of the subsequent stripes. Since the growth of a stripe 
depletes its surrounding, the source front will recover in concentration 
necessary for the emergence of a new stripe only above a certain distance from the old one. 
In order to simulate this scenario, in the $x\in(5,30)$ space units region of the 
simulation area a regular structure of tilted stripes with $c(x,y,0)=0$ were introduced. 
The angle between the edge of the simulation area (the $Y$-axis) and the stripes 
was about $30$ degrees, and the wavelength of the structure was about 
$8$ space units. When the front sweeps through this ``pre-patterned'' region, its
contribution will accumulate on the stripes with the unstable concentration $c=0$, 
leading to a fast phase separation. In this way, a stable striped structure, 
oblique to the front, will develop. However, an oblique striped structure
survives only around $v\approx 1$. At $v\approx 2$, slightly disturbed lamellar morphology formed. 
When $v>5$ and $v<0.1$, irregular morphology and spotty irregular morphology emerged in a $250$x$620$ 
simulation area. No significant change was observed when $c_f=\pm 0.12$.

\section{Rotating source fronts}

The oblique morphology in the $NaOH+CuCl_2$ reactions in PVA gel sheets usually 
appears when the traveling reaction-diffusion front changes its direction, while the newly 
forming stripes keep the orientation of previous stripes. 
This process was computationally modeled by a rotating source front segment in a simulation area of  
$1200$x$600$ space units, having a length of $570$ space units. The initial concentration in the whole 
simulation area was set to $c_0(x,y,0)=-1+\eta$.
The source term added to the Cahn-Hilliard equation takes the form

\begin{equation}
S(x,y,t;\omega)=
A\sqrt{(x-x_0)^2+(y-y_0)^2} \cdot 
\exp \left\{ -\alpha \left[(x-x_0) sin(\omega t+\theta_0)+(y-y_0) cos(\omega t+\theta_0)\right]^2\right\}
\end{equation}

The character of the pattern formation is a function of the front 
speed, that depends on the angular velocity, as well as the position along the radius.
In our simulations, having a front length of $570$ space units, striped patterns just behind the front 
appeared around the
angular velocity interval $\omega\in(0.002-0.02)$. The dynamics of the pattern formation was as follows:
The first stripe forms roughly along the initial position of the 
front. Although the orientation of the front is continuously altered, the newly formed stripes will 
``try'' to form along the old ones, parallel with them. Since the front changes 
its orientation in the meanwhile, the above scenario will lead to an oblique morphology.  

However, the simulations showed that the stripes cannot grow perpendicular to the front.
Their elongation becomes unstable when the angle of the stripes formed to the front supplying the
phase separating material reaches about $70-90$ degrees. At this stage, in some domains 
just behind the front oblique stripes with a small angle appear. In some other domains irregular 
morphology appears. Later, the above scenario may repeat itself. 

Note that in the vicinity of the outer
endpoint of the rotating front, where the speed is relatively high, the source front may not immediately 
be followed by the phase separation. The outer core of the circular region will be 
patterned by a different mechanism, namely the striping initiated by the arc-like edge where 
the concentration changes from $c=0$ to $c=-1$. This striping will start along the edge, and
will spread inside the unstable region, until it meets the straight striping initiated by the 
front itself.

Finally we review the pattern formation at much higher and lower angular velocities. 
When the angular velocity is higher than $\omega=0.05$, the overwhelming majority
of the phase separation takes place far behind the front. Two mechanisms play an 
important role in the pattern formation. As mentioned previously, a striping will be initiated by 
the arc-like edge, where the concentration changes from $c=0$ to $c=-1$. 
However, in the inner regions, mostly irregular patterns will form. 
In the case of low angular velocities, when $\omega<0.001$, spotty 
irregular morphology appears in the wake of the front with a length of $570$ space units.

\section{Quenching fronts revisited}

An alternative way to start the spinodal decomposition in the wake of a traveling front 
is to set the concentration of the phase separating compound $c$ in between the spinodal points, 
while the temperature is dropped below the critical value only behind the front. 
The quenching in our simulations has been realized by changing the sign of the second-order 
term in the Cahn-Hilliard equation. 

Pattern formation in the wake of quenching fronts has been studied by  
computational analyzes similar to those performed in the case of source fronts.
The main difference with respect to the pattern formation in the wake of the 
source fronts is the appearance of the columnar morphology at low front 
speeds, in agreement with Furukawa's results \cite{furukawa}. 
With the concentration of the phase separating compound set to $c_0(x,y,0)=0$, the columnar morphology 
appears below $v\approx 0.85$. At $v\approx 0.9-2$, more or less coarsed lamellar 
morphology forms. When $c_0(x,y,0)=\pm 0.1$, this limit where columnar morphology appears, 
shifts toward lower velocities. 

The formation of oblique morphologies has also been observed. 
The most extensive OM-s appeared around the velocity $v\approx 1$, when
oblique stripes with $c(x,y,0)=0.2$ and geometric parameters as above were initially 
introduced  in the $x\in(5,30)$ space units region of the simulation area. 

The most interesting characteristics of the pattern formation in a growing quenched area 
was found in the computational investigation of rotating fronts. Moving outwards along the
radius, that is, reaching front segments with higher velocities, columnar morphology, 
oblique patterns forming different angles to the front, and, 
near the border of the quenched region, lamellar morphologies have been observed. 

It is remarkable that the growing ends of the columnar stripes always remain perpendicular 
to the front (fig. 3). As a consequence, in the case of a front rotating with an appropriate speed, 
bent columnar structures build up, which develop into regular arcs.
This result can be of major importance in various nanotechnological processes, 
since it makes possible the ``wiring'' of a surface upon a previously given curve.
By moving the quenching front along an arbitrary non-intersecting path, a columnar structure 
of high and low density regions builds up behind it.

\section{Conclusion}

The formation of striped microscopic patterns behind a traveling reaction-diffusion 
front in the $NaOH+CuCl_2$ chemical system has been investigated.
The scenario of the pattern formation was modeled
by the terms of phase separation described by the Cahn-Hilliard Equation.
The formation of the stripes, being parallel or oblique to the reaction-diffusion front,
has been reproduced by computer simulations.

The results have been compared to the pattern formation in the case of directional quenching.
At low front velocities, the formation of striped patterns perpendicular to the quenching front 
have been observed, these patterns being absent in the case of the source fronts. 
In the wake of a slowly progressing front that simultaneously changes its direction, 
the growing endpoints of the stripes will always be perpendicular to the front. This effect 
enables one to ``draw'' on a surface regular stripes following an arbitrary curve. 
Such a patterning could be of major importance in nanotechnology.

\begin{acknowledgments}
We thank Zolt\'an R\'acz, Stoyan Gisbert, Nicholas R. Moloney, Gyula Bene, Emese No\'emi Sz\'asz 
and Istv\'an Forg\'acs for useful discussions. This work has been supported by the 
Sapientia-KPI Foundation,  Domus Hungarica 
Foundation of the Hungarian Academy of Sciences, and the Agora Foundation.  
\end{acknowledgments}


\newpage

\begin{figure}
\includegraphics{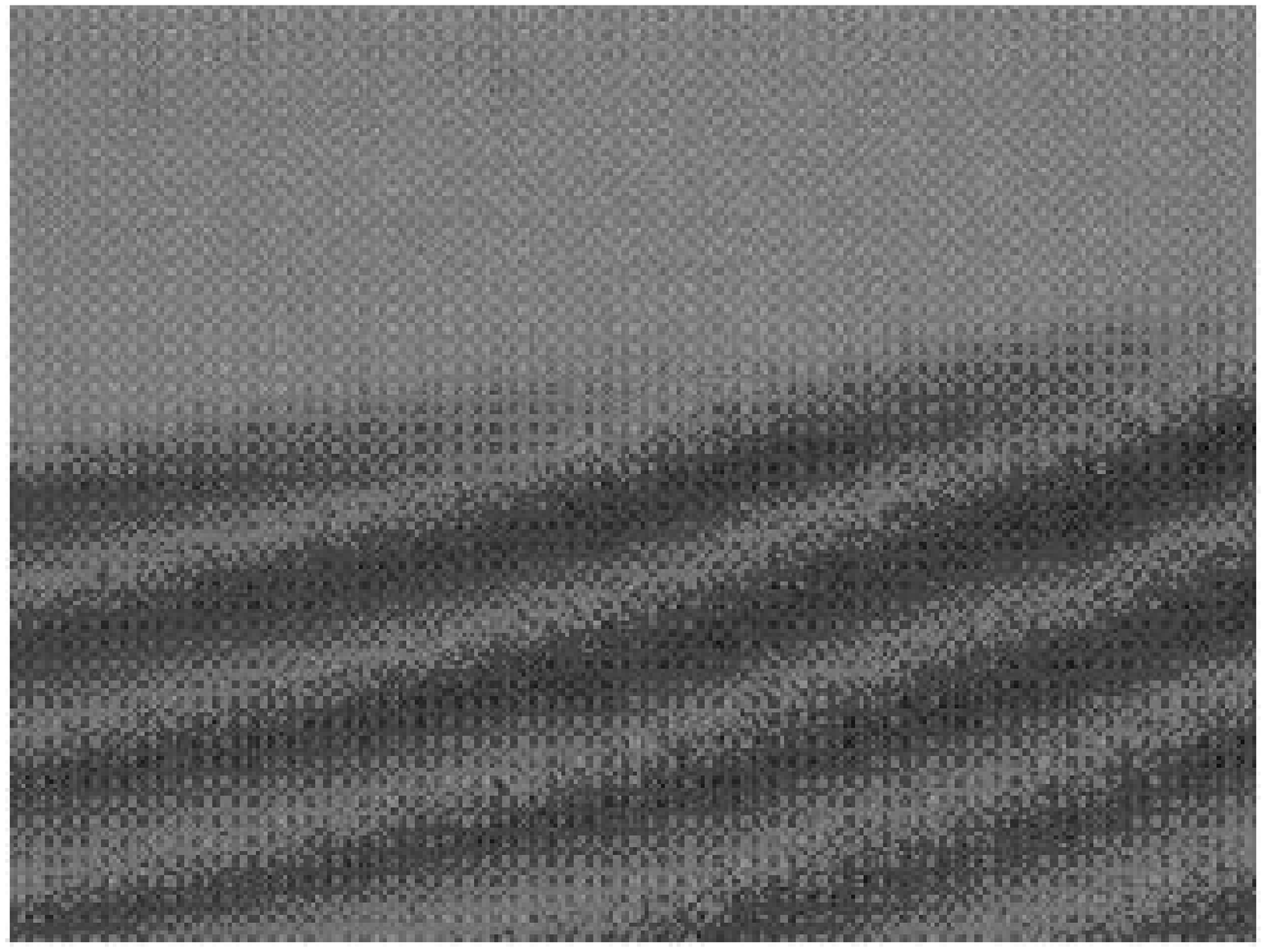}
\caption{\label{fig:exp} Stripes of colloidal $CuO$ grains of oblique morphology. The precipitate 
structures form as a result of $8\ M\ NaOH+ 0.586 \ M\ CuCl_2$ reaction in a thin PVA gel sheet 
of about $0.2\ mm$ thickness, located between a microscope slide and a cover glass. The width of the 
figure is $0.58\ mm$. The sharp straight line shortly below the endpoints of the stripes represents the 
active border of the system.}
\end{figure}

\begin{figure}
\includegraphics{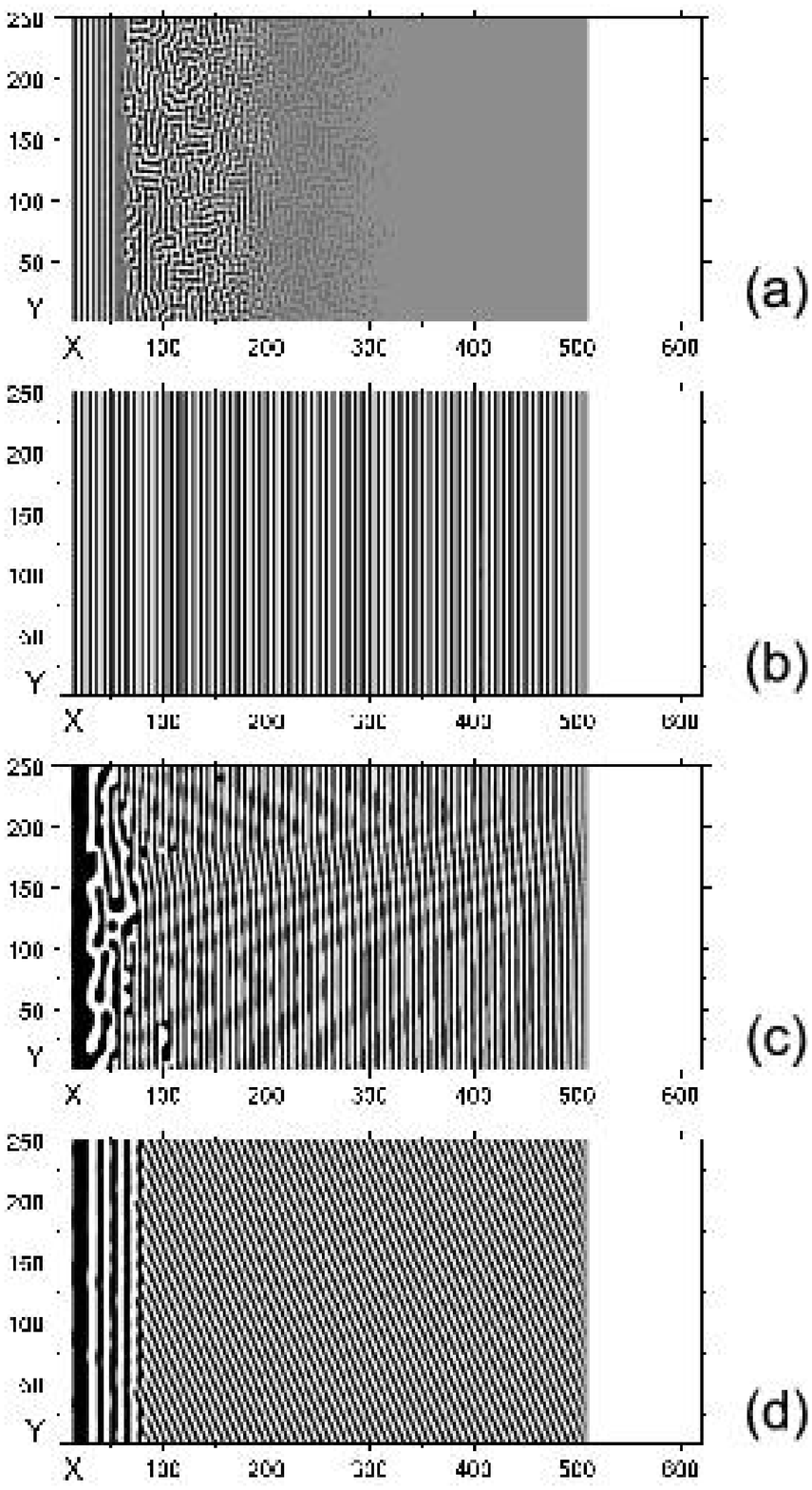}
\caption{
\label{fig:progr} Pattern formation in the wake of source fronts traveling with constant speed.
The dimensions of the simulation area are $250$x$620$ space units. $\epsilon=0.5$,
 $a=1$, $b=10$, $\alpha=0.1$  in all the sub-figures. A gray scale has been used, with white as $c=-1$ and 
black as $c=+1$.
(a.) Random patterns formed far behind to the front. Note the parallel striping that turns out in the 
wake of the random morphology.
$v=10$, $A=1.78$ ($c_f=0$), $t=50$.
(b.) Regular lamellar patterns.
$v=1$, $A=0.178$ ($c_f=0$), $t=500$. 
(c.) Slightly disturbed lamellar morphology and oblique morphology with a small tilting angle 
formed when the first stripes were destroyed by random spots located in the  $x\in(5,30)$ space 
units region.
$v=1$, $A=0.178$ ($c_f=0$), $t=500$.
(c.) Oblique morphology formed when the  $x\in(5,30)$  space units region was ``pre-patterned''
with tilted stripes forming an angle of about $30$ degrees with the $Y$-axis and having a wavelength of about 
$8$ space units.  Note the parallel striping that turns out in the 
wake of the oblique morphology.  $v=1$, $A=0.178$ $(c_f=0)$, $t=500$.
}
\end{figure}

\begin{figure}
\includegraphics{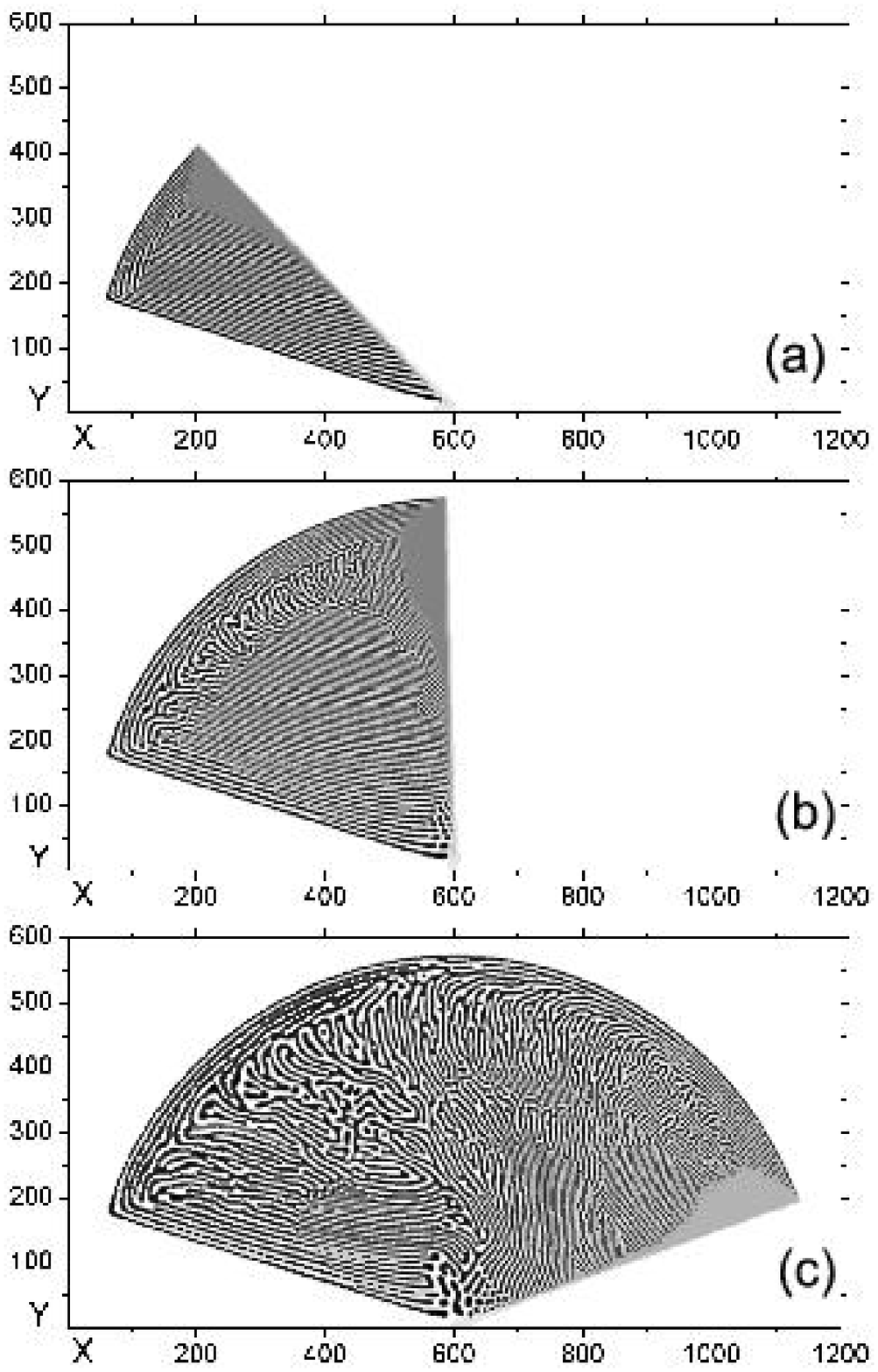}
\caption{\label{fig:rotchem} Pattern formation in the wake of rotating source fronts.
The angular velocity is $\omega=0.005$, the dimensions of the simulation area are
$1200$x$600$ space units. The parameter $\theta_0=0.3$ was introduced to prevent the first stripes 
being parallel to one of the grid lines. $\epsilon=0.5$, $x_0=600$, $y_0=5$, $A=0.00089$,
$\alpha=0.1$ ($c_f=1$), $c_0=-1$. Attention should be paid to the regions just behind the 
wake of the fronts; after this, coarsening will restructure the patterns.
(a.) Oblique striping at $t=100$.
(b.) The critical angle is reached, the growth of the stripes becomes unstable. $t=250$.
(c.) Oblique morphology with a new angle builds up in the wake of the front. $t=500$. 
}
\end{figure}

\begin{figure}
\includegraphics{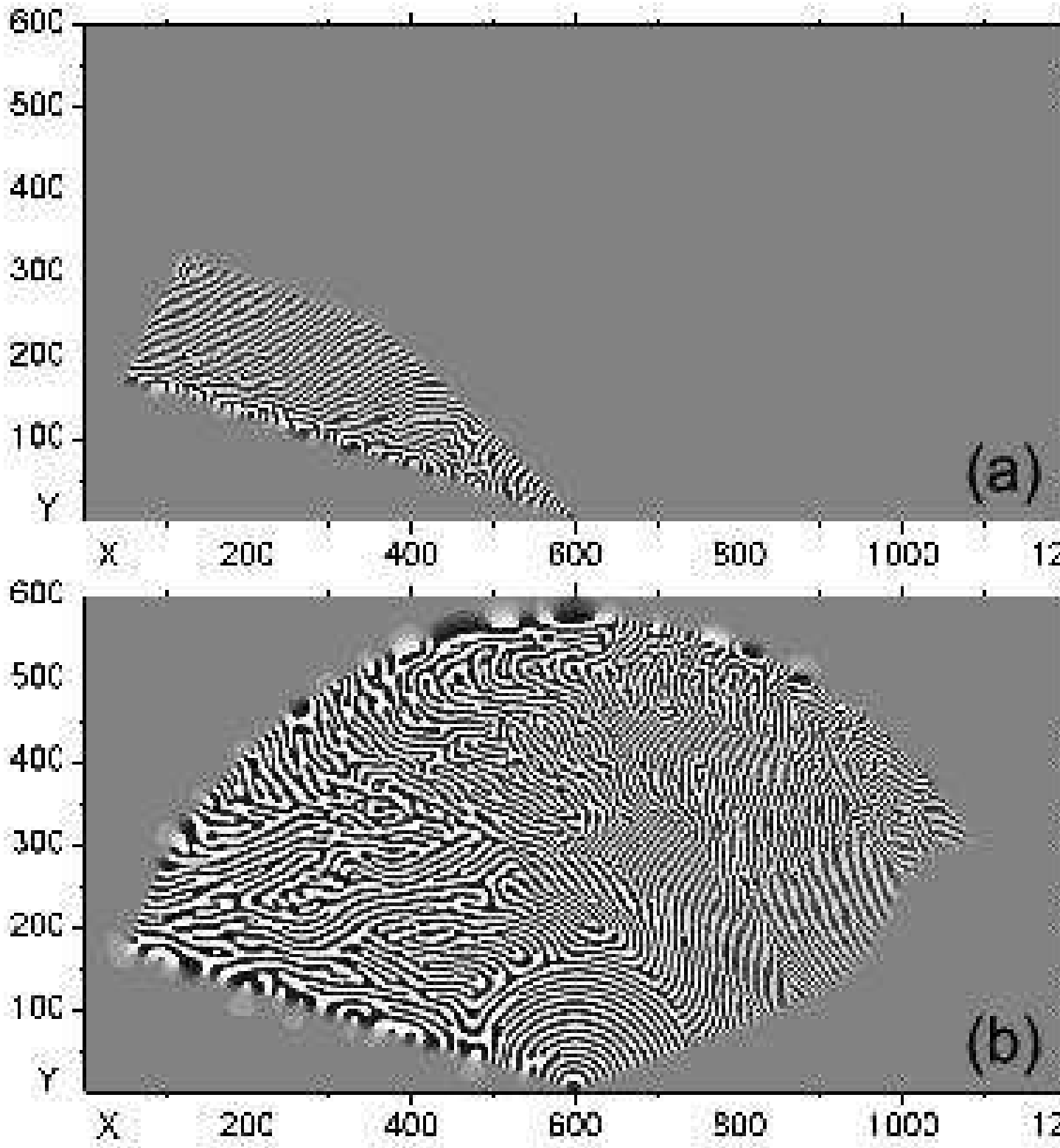}
\caption{\label{fig:rotq} Pattern formation in the wake of a rotating quenching front.
The dimensions of the simulation area are $600$x$1200$ space units. The parameter $\theta_0=0.3$ was 
introduced to prevent the stripes being parallel to the grid lines. $\omega=0.005$,
$\epsilon=0.5$, $x_0=600$, $y_0=5$, $c_0=0$; $t=100$ (a.) and $t=500$ (b.). 
Note the continuous arc-shaped stripes in the vicinity of 
the rotation center, where the local velocity is small. Attention should be paid to the regions 
just behind the front; after this, coarsening will restructure 
the patterns, except for the arcs ``pulled'' by the front.}
\end{figure}

\end{document}